\documentclass[%
aip,
amsmath,amssymb,
reprint,%
]{revtex4-1}

\usepackage{graphicx}
\usepackage{dcolumn}
\usepackage{bm}

\usepackage[utf8]{inputenc}
\usepackage[T1]{fontenc}
\usepackage{mathptmx}
\usepackage{etoolbox}
\usepackage{amsmath}
\usepackage{amssymb}
\usepackage[version=4]{mhchem}
\usepackage{textcomp}
\usepackage{color}
\usepackage[colorlinks=true, letterpaper=true, pdfstartview=FitV, linkcolor=blue, citecolor=blue, urlcolor=blue]{hyperref}
\usepackage{lineno}
\usepackage{xcolor}
\makeatletter
\def\@email#1#2{%
	\endgroup
	\patchcmd{\titleblock@produce}
	{\frontmatter@RRAPformat}
	{\frontmatter@RRAPformat{\produce@RRAP{*#1\href{mailto:#2}{#2}}}\frontmatter@RRAPformat}
	{}{}
}%
\makeatother
\begin{document}


\title{Engineering 2D high-temperature ferromagnets with large in-plane anisotropy via alkali-metal decoration in a tetragonal CoSe monolayer}


\author{Yiran Peng}
\author{Yanfeng Ge}
\author{Yong Liu}
\author{Wenhui Wan}
\email{wwh@ysu.edu.cn}
\affiliation{State Key Laboratory of Metastable Materials Science and Technology $\&$ Hebei Key Laboratory of Microstructural Material Physics, School of Science, Yanshan University, Qinhuangdao 066004, P. R. China.}
\begin{abstract}
Two-dimensional (2D) ferromagnetic materials with high Curie temperature ($T_{\rm c}$) and large magnetic anisotropy energy (MAE) are critical for nanoscale spintronics but remain rare. We propose, via first-principles calculations, that adsorbing alkali atoms ($A$ = Li, Na, K, Rb, Cs) onto a tetragonal CoSe monolayer transforms it into a series of stable 2D ferromagnetic metals, $A$CoSe, with an in-plane easy axis. Notably, LiCoSe is a half-metal. These functionalized monolayers exhibit dramatically enhanced ferromagnetism compared to the pristine layer, with $T_{\rm c}$ > 300 K and MAE > 800 $\mu$eV/Co. 
The coupled alkali atoms amplify the local magnetic moment of Co ions, reinforce ferromagnetic Ruderman–Kittel–Kasuya–Yosida (RKKY) and superexchange couplings, and concurrently weaken the direct antiferromagnetic exchange between Co ions. Furthermore, tensile strain can further elevate the MAE (via band shifting) and increase $T_{\rm c}$ (by strengthening the nearest-neighbor exchange $J_1$). 
Among them, NaCoSe exhibits the highest MAE and excellent strain-modulated $T_{\rm c}$, rendering it the most promising candidate material.
Our results establish alkali-metal decoration as an effective strategy for realizing 2D ferromagnets with high $T_{\rm c}$ and large MAE in tetragonal lattices.
\end{abstract}


\maketitle

Spintronic devices are widely studied for their high-speed operation, nonvolatility, high storage density, low power consumption, and compact size.\cite{Jia2025} A major advance came in 2017 with the discovery of long-range ferromagnetic (FM) order in 2D CrI$_3$\cite{Huang2017} and  Cr$_2$Ge$_2$Te$_6$,\cite{Gong2017} sparking intense interest in 2D magnets. For practical use, room-temperature ferromagnetism and larger magnetic anisotropy energy (MAE) is essential.\cite{Guo2021} Several high-temperature 2D ferromagnets have been synthesized---including VSe$_2$,\cite{Bonillaa2018,Yu2019} FeS, \cite{Zhang2023} CrTe$_2$,\cite{Wu2021} MnSe$_2$ \cite{O’Hara2018}, CuCr$_2$Se$_4$ \cite{Zhang2022} and Fe$_3$GaTe$_2$\cite{a1}---with Curie temperatures ($T\rm_c$) reaching 300--470  K.\cite{Yu2019,Zhang2023,Wu2021,O’Hara2018,Zhang2022,a1}
However, the number of such high-performance 2D FM materials remain scarce. 
Designing and synthesizing 2D ferromagnets with high $T\rm_c$ and large MAE remains a key challenge for next-generation spintronics.

The exceptional strain tolerance and high surface-to-volume ratio of 2D materials render their physical properties highly tunable. These magnetic properties, particularly the $T_{\rm c}$ and MAEs, can be effectively modulated through various external perturbations, such as strain, defect introduction, substrate coupling, carrier doping, and surface absorption.\cite{Wu2023,Ma2020,Sun2023,D0NR04663A,Verzhbitskiy2020,Shen2021}
Recent advances in gate-controlled ion injection technology \cite{Miyagawa2024} have enabled precise intercalation or surface deposition of alkali atoms in thin-film systems.
Experimentally, Deng et al. demonstrated that the electron doping from lithium (Li) intercalation enhance the itinerant ferromagnetism in Fe$_3$GeTe$_2$ thin flakes and sharply elevating $T_{\rm c}$ to room temperature.\cite{Deng2018}
Ma et al. realized a phase transition between superconductor and FM insulating states in (Li, Fe)OHFeSe thin flakes through gate-driven Li intercalation. \cite{Ma2019} 
Through first-principles calculations, Khan et al. predicted that the decoration of a CrSnSe$_3$ monolayer with alkali metals (Li, Na, K) can raise $T_{\rm c}$ from 51 K to 241--265 K.\cite{Khan2023}
Collectively, these studies establish alkalimetal decoration as a viable route to modulate the magnetism of 2D materials.

Here, we focus on 2D tetragonal materials $MX$ ($M$ = transition metal, $X$ = nonmetal), comprising of 166 structurally stable members.\cite{Xuan2022}
These materials feature hollow sites above the nonmetal atoms, providing ideal adsorption sites for alkali atoms and offering a platform for studying how alkali-metal decoration influences their magnetic properties. 
Among these materials, the CoSe monolayer in the $P4/nmm$ phase has been synthesized \cite{Zhou2016,Liu2018,Shen2018} and exhibits weak ferromagnetism with $T_c$ of only 8 K.\cite{Peng2025}
Meanwhile, the bulk phases of $A$Co$_2$Se$_2$ ($A$ = K, Rb, Cs) crystallize in the ThCr$_2$Si$_2$-type structure, consisting of planar FM sheets that are either aligned (for KCo$_2$Se$_2$ and RbCo$_2$Se$_2$) or antialigned
(for CsCo$_2$Se$_2$).\cite{Yang2013,Liu2015,Wdowik2015}
These findings suggest that $A$CoSe ($A$ = alkali metal) monolayers could potentially be obtained either by decorating CoSe monolayer with alkali metals or by exfoliating their bulk counterparts.  
However, the magnetic properties of alkali-metal decorated CoSe monolayer have not yet been systematically studied, limiting their potential applications in spintronics.

In this work, through first-principles calculations, we have confirmed that $A$CoSe ($A$ = Li, Na, K, Rb, Cs) monolayers are stable FM metals. Especially, the LiCoSe monolayer is a half-metal. 
Compared to the CoSe monolayer, all five $A$CoSe monolayers exhibit room-temperature ferromagnetism and large in-plane MAEs.
The coupling of alkali atoms with CoSe sheet increases the magnetic moments of Co ions, strengthens the ferromagnetic RKKY coupling, and weakens direct antiferromagnetic (AFM) direct coupling in the CoSe system.
Moreover, applying tensile strain can further increase both the MAE and $T_{\rm c}$. These results indicates that CoSe monolayer with alkali metal-decoration are high-$T_{\rm c}$ FM materials, desirable for spintronic applications.

First-principles calculations were performed with the Vienna Ab initio Simulation Package (VASP),~\cite{Kresse1996} using the projector augmented wave (PAW)~\cite{Bloechl1994} pseudopotentials and the Perdew, Burke, and Ernzerhof (PBE)~\cite{Perdew1996} exchange-correlation functionals.
A 20 {\AA} vacuum layer was added along the $z$ direction to avoid the interaction between periodic images. We employed a kinetic energy cutoff of 400 eV and a $14 \times 14 \times1$ k-point mesh for Brillouin zone sampling. The convergence criteria for force and energy were $10^{-2}$ eV/{\AA} and $10^{-6}$ eV, respectively. 
We estimated the MAE with a dense k-mesh of $30\times 30\times 1$ with the spin–orbit coupling (SOC) effect being considered.
The phonon spectrum were obtained through the finite displacement method as implemented in the Phonopy code.\cite{Togo2015} 
$Ab$ $initio$ molecular dynamics (AIMD) simulations were carried out for 5 ps at 300 K, using a $4\times 4\times 1$ supercell and canonical (NVT) ensemble.\cite{Martyna1992}
The Curie temperature ($T\rm_c$) was calculated via Monte Carlo simulations based on a $64\times 64\times 1$ supercell using the MCsolver package.\cite{Liu2019}

Based on available experiments,\cite{Yang2013} we examined the electronic correlation effect in bulk K(CoSe)$_2$ and Rb(CoSe)$_2$ by varying effective on-site Coulomb interactions ($U\rm _{eff}$)\cite{Dudarev1998} for Co-$3d$ electrons. Figs. S1 and S2 show that only small $U\rm _{eff}$ reproduces the observed FM ground state in bulk phase.\cite{Yang2013} Furthermore, zero $U\rm _{eff}$ yields magnetic moments and lattice constants closest to measured values. Thus, we performed all subsequent calculations of alkali-metal decorated CoSe monolayer without an explicit on-site Coulomb correction ($U\rm _{eff} = 0$ eV).

\begin{figure}[tbp!]
	\centerline{\includegraphics[width=0.45\textwidth]{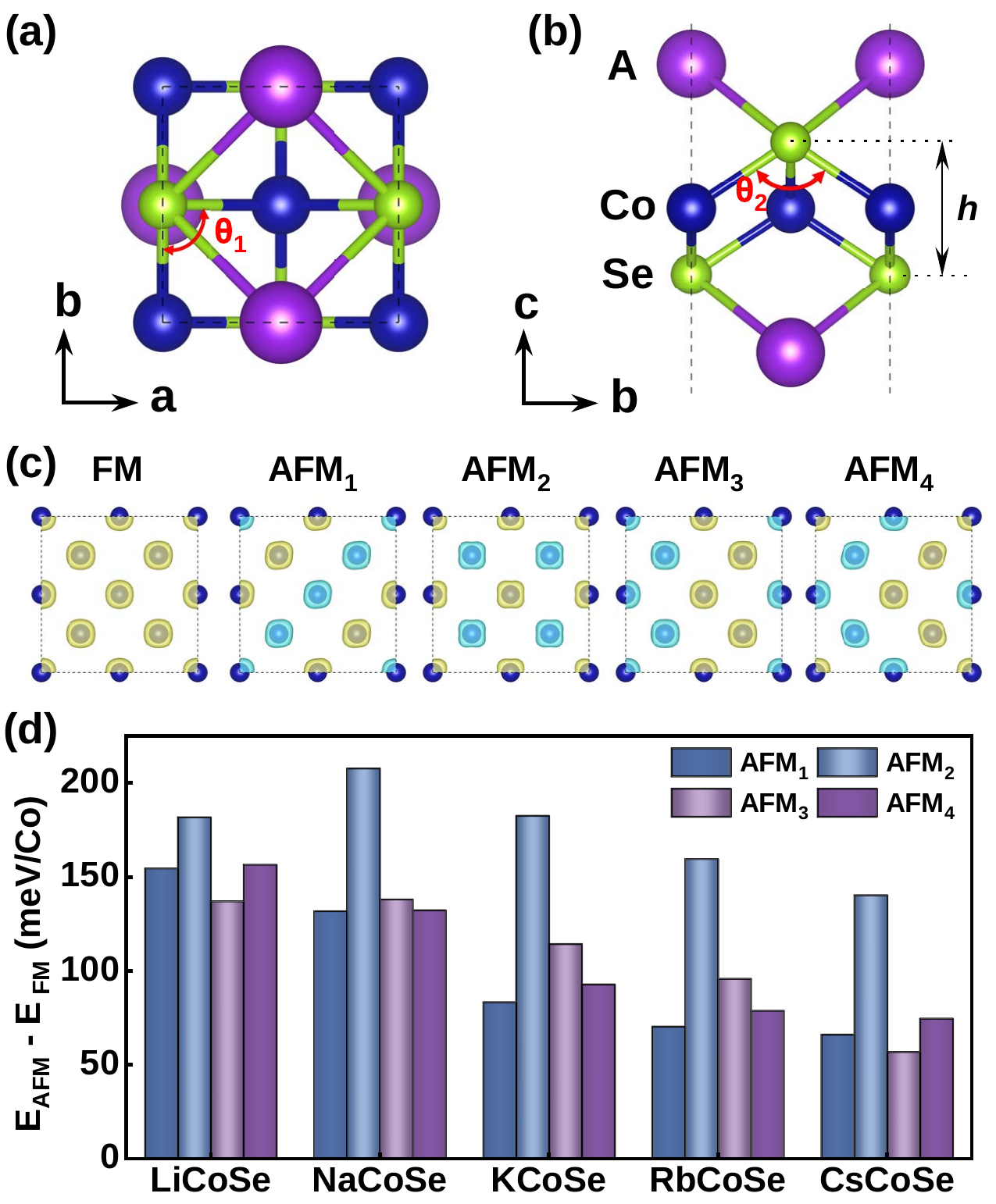}}
	\caption{(a) The top view and (b) side view of the $A$CoSe ($A$ = Li, Na, K, Rb, Cs) monolayers. (c) The spin configuration of the FM state and four different AFM states. Yellow and blue color represent the up and down spins, respectively. (d) The energies of AFM states relative to FM state.}
	\label{fg1}
\end{figure}

The adsorption of alkali metal ($A$ = Li, Na, K, Rb, Cs) atoms on the both sides of CoSe monolayer forms the five-layer $A$CoSe structure with an $A$-Se-Co-Se-$A$ stacking sequence. 
Energy comparisons of various adsorption sites [Fig.~S3] confirm the most stable configuration to be the hollow site above the Se atoms on the opposite side. The configuration with full occupation of hollow site (i.e., 100\% coverage) yields the lowest formation energy [Fig.~S3(f)].
The resulting $A$CoSe monolayer [Fig.~\ref{fg1}(a, b)] exhibits a typical anti-MXene structure\cite{La2025} with the space group $P4/nmm$ and the point group $D_{4h}$. Each Co atom bonds with four Se atoms, while each Se atom bonds with four Co atoms and four $A$ atoms. Compared with the intrinsic CoSe, alkali-metal adsorption enlarges the lattice constant ($a$) and the Co-Se bond lengths ($d_{\rm Co-Se}$) [see Table~\ref {tab:table1}]. 
The vertical distance ($h$) between Se layer [Fig.~\ref{fg1}(b)] initially rises and subsequently decreases as alkali metal varies from Li ato Cs.
Correspondingly, the Co-Se-Co bond angles ($\theta_1$, $\theta_2$) first decrease then increase along the same series.

To determine the magnetic ground state, we considered the FM state and four different AFM states for each $A$CoSe monolayer [Fig.~\ref{fg1}(c)]. Their energies are shown in Fig.~\ref{fg1}(d). All five $A$CoSe ($A$ = Li, Na, K, Rb, Cs) monolayers adopt the FM state as their magnetic ground states.   

The $A$CoSe monolayer in FM state exhibits excellent structural stability. The formation energy for the absorption process is negative [Fig.~S3(e)], indicating that the synthesis pathway of $A$CoSe is exothermic and feasible. 
The dynamics stability of $A$CoSe is verified by their phonon spectrum [Fig. S4] which exhibits positive frequencies across the Brillouin zone. MD simulations at 300 K shows that $A$CoSe exhibits minor lattice distortion, indicating their thermodynamic stability [Fig.~S4]. Furthermore, their elastic constants [Table S1] satisfy the Born criteria for square lattice: $C_{11} > 0$, $C_{66} > 0$, and $|C_{11}| > |C_{12}|$,\cite{Mazdziarz2019} exhibiting their mechanical stability. 
Finally, considering the high reactivity of alkali metals, we proposed encapsulating $A$CoSe with an insulating $h$-BN to protect its magnetic properties under ambient conditions [Fig.~S5].

\begin{table}[b!]
	\caption{\label{tab:table1}	Lattice constant ($a$), Co-Se bond length ($d_{\rm {Co-Se}}$), Co-Se-Co bond angles ($\theta_1$ and $\theta_2$), vertical height ($h$) between two Se layers in intrinsic and alkali-metal-decorated CoSe monolayers.}
	\begin{ruledtabular}
		\centering %
		\begin{tabular*}{0.45\textwidth}{@{\extracolsep{\fill}}ccccccc}
			& $a$  & $d_{\rm {Co-Se}}$  &   $\theta_1$ &$\theta_2$&$h$\\
			&    (\AA)& (\AA) & (\textdegree) &(\textdegree)&(\AA) \\
			\hline
			CoSe        &3.733&2.307&69.80&108.02&2.711\\
			LiCoSe      &3.763&2.463&65.33&99.51&3.185\\
			NaCoSe      &3.927&2.448&69.12&106.68&2.923\\
			KCoSe       &4.013&2.411&72.11&112.69&2.671\\
			RbCoSe      &4.049&2.406&73.02&114.58&2.593\\
			CsCoSe      &4.102&2.410&73.98&116.61&2.532\\
		\end{tabular*}
	\end{ruledtabular}
\end{table}

\begin{figure}[tbp!]
	\centerline{\includegraphics[width=0.45\textwidth]{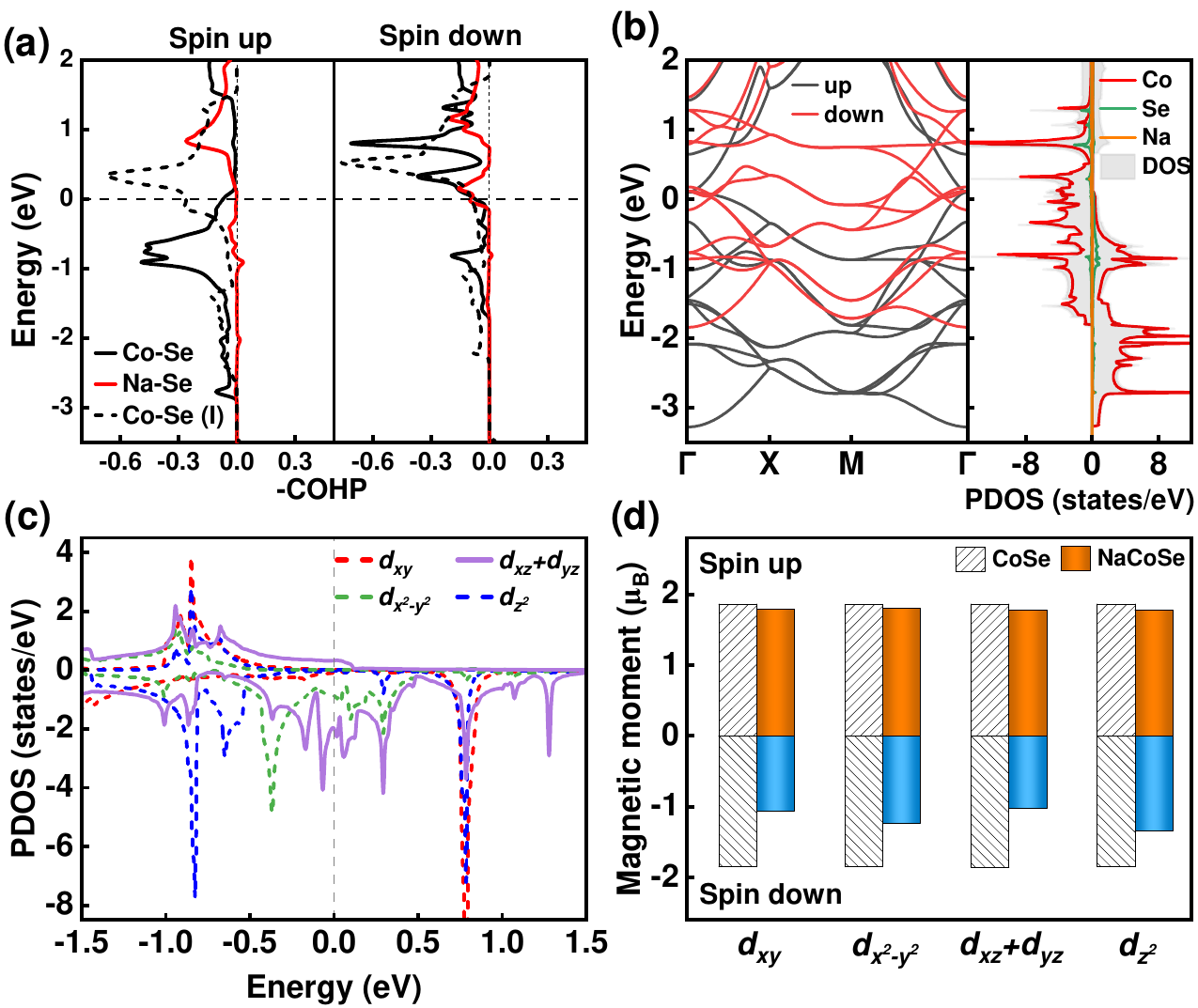}}
	\caption{(a) The -COHP values of Co-Se and Na-Se bonding in spin-up and spin-down channels for NaCoSe monolayer. For comparison, dashed lines (Co-Se(I)) is the Co-Se bonds in intrinsic CoSe monolayer. (b) The band structure and (c) the contribution of Co-$d$ orbitals to the density of states. (d) The contribution of Co-$d$ orbitals to the magnetic moment in NaCoSe and CoSe monolayers.}
	\label{fg2}
\end{figure}

Alkali-metal adsorption strengthens the Co–Se bonding in CoSe monolayer. Negative Crystal Orbital Hamilton Population (-COHP)\cite{Deringer2011} is employed to analyze the chemical bonding, where positive and negative values correspond to bonding and anti-bonding character, respectively. Fig.~\ref{fg2}(a) shows that the Co–Se bond in the intrinsic CoSe monolayer exhibits anti-bonding character near the Fermi level. In contrast, the adsorption of alkali metals significantly reduces this anti-bonding character [Fig.~\ref{fg2}(a) and Fig.~S7]. 
Additionally, alkali-metal adsorption form a Se–$A$ bond with weak bonding character. Consequently, the Young’s modulus ($Y$) of $A$CoSe becomes larger than that of the CoSe [Table S1].
$A$CoSe monolayer exhibits anisotropic in-plane $Y$ [Fig.~S6], with maximum value along the $x$ or $y$ direction and minimum one along the diagonal ($45^\circ$) direction. The $Y$ of $A$CoSe monolayers is lower than that of graphene (345 N/m)~\cite{Kudin2001} and \ce{MoS_2}  (118 N/m),~\cite{Cooper2013} indicating its good flexibility. 

Electronic structure calculations [Fig.~\ref{fg2}(b) and Fig.~S8] identify LiCoSe as a half-metal, with the other $A$CoSe being magnetic metals. 
We verified the electronic structure using the more accurate HSE06 functional, which gave consistent results [Fig. S8]. The densities of states (DOS) near the Fermi level ($E_F$) is dominated by the Co-$3d$ orbitals and Se-$4p$ orbitals, primarily from spin-down $d_{xz}+d_{yz}$ and $d_{x^2-y^2}$ states. Only LiCoSe has a band gap in the spin-up channel, while this gap is closed in other four $A$CoSe monolayers [Fig.~S9]. 
Meanwhile, compared to intrinsic CoSe, alkali metal decoration increases the DOS near the Fermi level ($D(E_F)$). The similar phenomena have been reported in alkali-metal decorated Cr$_2$Ge$_2$Te$_6$\cite{Song2019} and MoSi$_2$N$_4$.\cite{Sun2022}
We calculated the spin polarization rare $P = \frac{D_{\downarrow} - D_{\uparrow}}{D_{\downarrow} + D_{\uparrow}} \times 100\%$, where $D_{\uparrow}$ and $D_{\downarrow}$ represent the $D(E_F)$ in the spin-up and spin-down channels, respectively. The $P$ reaches 100\% in LiCoSe, while 68.63\%, 70.15\%, 66.7\%, and 53.86\% for Na-, K-, Rb-, and CsCoSe, respectively. These value are higher than intrinsic CoSe (39.9\%)\cite{Peng2025} and comparable to Fe$_2$H (77.8\%) \cite{Ding2022} and MnGaN (79\%).\cite{Ma2020}

The electronic properties of $A$CoSe are robust to strain. We defined the strain as $\varepsilon = (a/a_0-1) \times 100\%$, where $a$ and $a_0$ are the strained and intrinsic lattice constants, respectively. 
Fig.~S8 demonstrates that LiCoSe remains a half metal under strains ranging from -4\% to 4\%, whereas the other $A$CoSe compounds are magnetic metals throughout this range.

The alkali-metal adsorption enhances the asymmetry in the occupation of Co-$3d$ orbitals [Fig.~\ref{fg2}(d) and Fig. S10]. In intrinsic CoSe, the near-symmetric occupation of the Co-$3d$ orbitals results in a small magnetic moment of 0.403 $\mu_B$/Co.
In contrast, alkali-metal adsorption make some of the locally localized $3d$ electrons transform into bonding electrons [Fig.~\ref{fg2}(a)].
The net magnetic moment of Co ions in $A$CoSe ($A$ = Li, Na, K, Rb, Cs) increases substantially to 1.727, 1.689, 1.520, 1.439, and 1.458 $\mu_B$/Co, respectively.
Besides, Se ions acquire small induced magnetic moments due to proximity polarization. 

We applied the Stoner criterion, $I \times D(E_F) > 1$, to explain the emergence of ferromagnetism in the $A$CoSe monolayer. The Stoner parameter $I$ is evaluated as $I = \langle \varepsilon_{\rm k} \rangle / m_{\rm avg}$, where $\langle \varepsilon_{\rm k} \rangle$ and $m_{\rm avg}$ denote the exchange splitting energy and the average magnetic moment, respectively. Alkali‑metal adsorption increases both magnetic moments and $D(E_F)$, resulting in $I \times D(E_F)>1$ for all $A$CoSe monolayers [Table~S2], confirming the FM stability over the paramagnetic state. 

The Bader charge analysis shows charge transfer from the alkali atoms to the neighboring Se atoms [see Table S3]. These conduction electrons enhance the itinerant ferromagnetism through the Ruderman-Kittel-Kasuya-Yoshida (RKKY) exchange interaction.\cite{R1,K1,Y1} Meanwhile, the enlarged Co–Co distance weakens direct antiferromagnetic exchange, while the increase in $\theta_1$ toward 90$^{\circ}$ promotes ferromagnetic superexchange between the Co ions [Table~\ref{tab:table1}]. Together, these factors enhance the ferromagnetism in the $A$CoSe ($A$ = Li, Na, K, Rb, Cs) monolayers compared to intrinsic CoSe.

\begin{figure}[tbp!]
	\centerline{\includegraphics[width=0.5\textwidth]{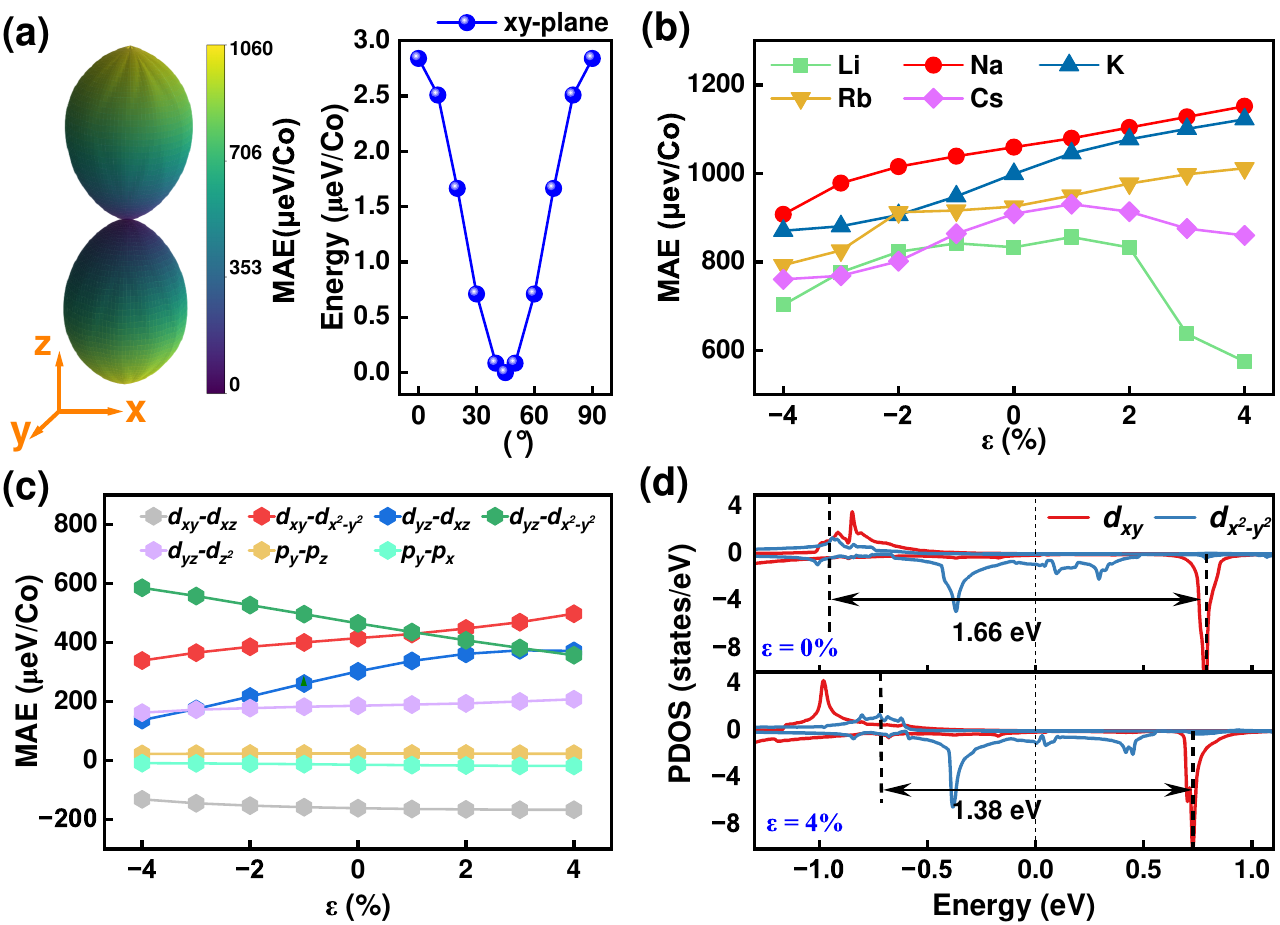}}
	\caption{(a) The energy of NaCoSe monolayer with spin direction lying on the whole space and $xy$ plane. (b) The strain-dependent MAE of $A$CoSe ($A$ = Li, Na, K, Rb, Cs) monolayers. (c) The contribution from the different orbital hybridization to the MAE of NaCoSe under strains. (d) The DOS of the $d_{xy}$ and $d_{x^2-y^2}$ states of NaCoSe under strains ($\varepsilon$) of 0\% and 4\%.}
	\label{fg3}
\end{figure}

A large MAE stabilizes the magnetization against thermal (spin) fluctuations.\cite{Yang2022} We calculated the total energy of $A$CoSe ($A$ = Li, Na, K, Rb, Cs) monolayers with different spin orientations.
All the $A$CoSe monolayer have an in-plane magnetization easy axis, which is along the [110] direction, corresponding to azimuth angle of $\theta = 90^{\circ}$ and $\varphi = 45^{\circ}$ in the Spherical coordinates. The MAE is estimated as $E\rm_{MAE}$ = $E_{[001]}$-$E_{[110]}$, whose values are 833, 1059, 998, 925, and 909 $\mu$eV/Co atom for $A$CoSe ($A$ = Li, Na, K, Rb, Cs) monolayers, respectively. The Co atoms dominate MAE. These MAEs are about 15 times higher than that of intrinsic CoSe (65.5 $\mu$eV/Co) \cite{Peng2025}, and other alkali-metal decorated materials, such as Li$_2$Fe$_2$SSe (172 $\mu$eV/Fe) \cite{Guo2021a} and NaCrTe$_2$ (460 $\mu$eV/Cr).\cite{Xu2020} 
Based on the classic magnetic dipole-dipole interaction [Part 7 of the Supplementary Materials], we estimated the contribution of magnetic shape anisotropy (MSA)\cite{Johnston2016} to the MAE. This contribution is only 26.63--57.22 $\mu$eV/Co atom---much smaller than that of the magnetocrystalline anisotropy---and is therefore neglected in the following analysis of MAE.

Tensile strain can further enhance the MAEs of $A$CoSe monolayers, except for LiCoSe [Fig.~\ref{fg3}(b)]. 
We focus on NaCoSe due to the largest MAE among $A$CoSe monolayers. The phonon calculations ensure its lattice stability under strains [Fig.~S12]. When tensile strain reach 4\%, its MAE increases to 1151 $\mu$eV/Co. To explain that, we expressed the MAE based on the second-order perturbation theory\cite{Wang1993}
\begin{small}
	\begin{align} \label{MAE} 
		{\rm MAE} &= E_{[001]} - E_{[110]} \nonumber\\
		&= \xi^2 (1-2\delta_{\alpha\beta})  \sum_{o,\alpha,u,\beta} \frac{
		|\left \langle {{o^\alpha}} \right|{\hat L_z}\left| {u^\beta} \right\rangle |^2-
			|\left\langle {{o^\alpha}} \right|{ \frac{\hat L_x + \hat L_y}{\sqrt{2}}} \left| {u^\beta} \right\rangle |^2}{\varepsilon^{\alpha}_{u}-\varepsilon^{\beta}_{o}}.
	\end{align}
\end{small}
where $\xi$, $\hat L_{z(x)}$, $\varepsilon_{o}$, and $\varepsilon_{u}$ are the SOC strength, the angular momentum operators, the energy levels of occupied states and unoccupied states, respectively.
$\alpha$ and $\beta$ are the spin index "+" and "-". $\delta_{\alpha\beta}$ is the Kronecker symbol. 
Considering the denominator ${\varepsilon_{u}-\varepsilon_{o}}$ of Eq.\ref{MAE}, the electronic states near the $E_F$ dominate the MAE.
Fig.~\ref{fg3}(c) shows the contribution of various orbital hybridization to the MAE under different strains. The increase in the MAE of NaCoSe is mainly due to the increased positive contribution from the hybridization $\langle d_{xy} |\hat L_z|  d_{x^2-y^2}\rangle$ and $\langle d_{xz} |\hat L_x|  d_{yz}\rangle$ between different spin channels. Fig.~\ref{fg3}(d) and Fig.~S13 shows that tensile strain reduces the energy difference between the corresponding DOS peaks near the Fermi level, decreasing the denominator in Eq.~\ref{MAE} and raising the MAE.

\begin{figure}[tbp!]
	\centerline{\includegraphics[width=0.45\textwidth]{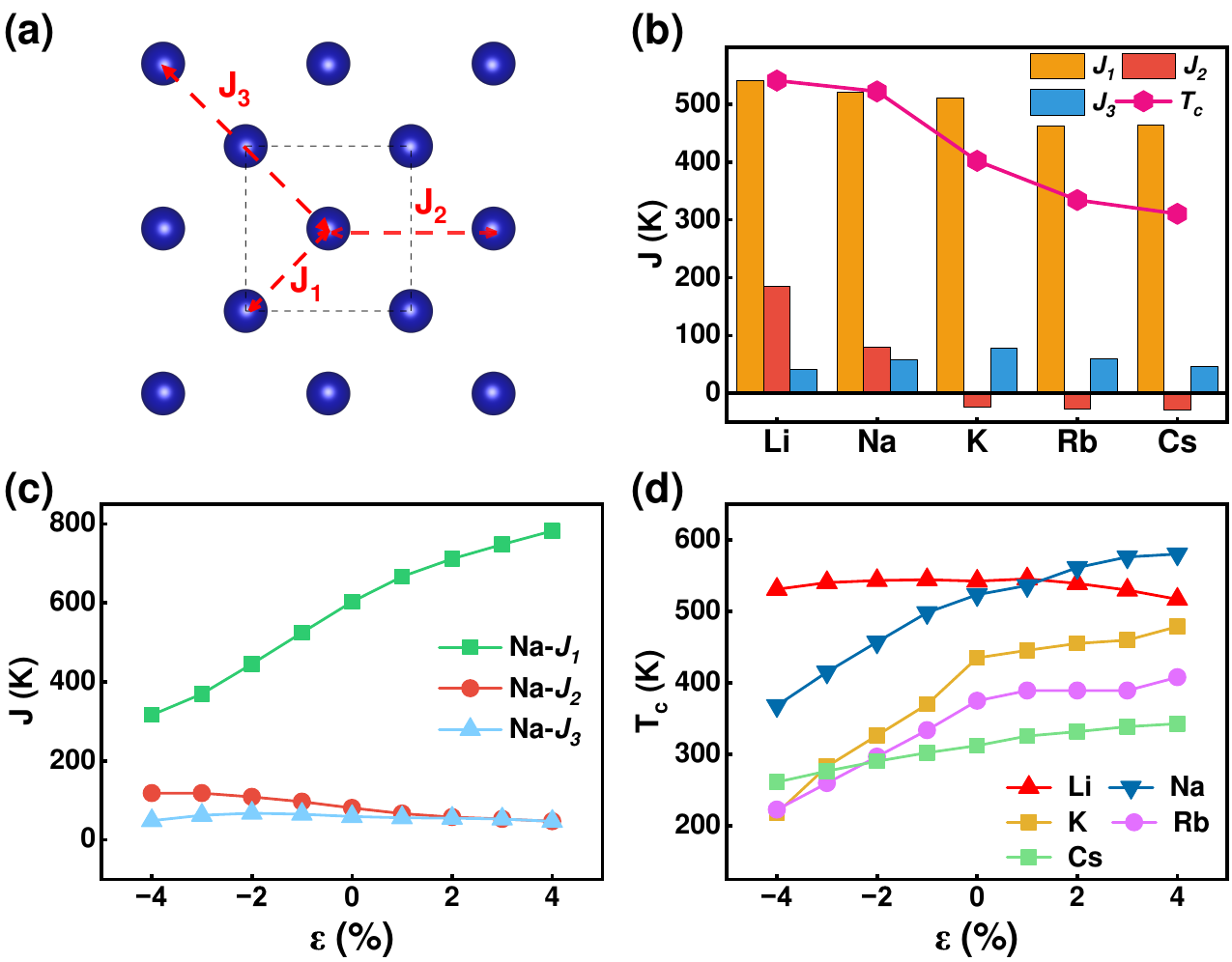}}
	\caption{(a) The exchange coupling constants $J_1$, $J_2$, $J_3$ between the nearest-neighboring, next-nearest-neighboring, and next-next-nearest-neighboring Co atoms. (b) The exchange coupling constants and $T\rm _c$ in strain-free $A$CoSe ($A$ = Li, Na, K, Rb, Cs) monolayer. (c) The exchange coupling constants of NaCoSe monolayer under different strains. (d) The $T\rm _c$ of $A$CoSe varying with strains. }
	\label{fg4}
\end{figure}

To calculate the $T\rm_c$, we used the Heisenberg model to describe the Hamiltonian: 
\begin{eqnarray}
	H=-\sum_{i,j}J_{ij}\boldsymbol{S}_i\cdot \boldsymbol{S}_j-D\sum_i\left(S_i^{e}\right)^2,
	\label{heise}
\end{eqnarray}
where $J_{i j}$ is the exchange constant and $D$ is the MAE.
We considered $J_{i j}$ up to the third-nearest neighbors, including $J_1$, $J_2$, and $J_3$ [Fig.~\ref{fg4}(a)]. $\boldsymbol{S}_i$ is the spin vector on the $i$-th site and $S_i^{e}$ the spin vector on the easy axis. With spin being normalized the to 1, we expressed the total energies per Co atom of different magnetic states, as shown in Fig.~\ref{fg1}(c), as
\begin{eqnarray}
	E_{\rm FM} &=&E_{0}- 2J_1 - 2J_2 - 2J_3 - D,\\
	E_{\rm AFM_1}&=&E_0 + 2 J_2 - 2 J_3 -D,\\
	E_{\rm AFM_2}&=&E_0 + 2 J_1 - 2 J_2 - 2 J_3 -D,\\
	E_{\rm AFM_3}&=&E_{0}+ 2 J_3  - D, \\
	E_{\rm AFM_4}&=&E_{0}+ 2 J_2- D.
\end{eqnarray}
We substituted the energies from the first-principles calculations into above equations and got the exchange constants, as summarized in Fig.~\ref{fg4}(b). The nearest-neighbor exchange constants $J_1$ dominates magnetic coupling between Co ions. 
The MC simulations predicted that the $T_{\rm c}$ of $A$CoSe monolayers exceed room temperature [Fig.~\ref{fg4}(b) and Fig.~S11], 
which is larger than that of CoSe (8K)\cite{Peng2025} and other alkali-decorated systems including NaCrSe$_2$ (226 K)\cite{Xu2020} and NaCrSnSe$_3$ (256 K).\cite{Khan2023}

The adsorbed alkali atoms induce lattice expansion and extra coupling with CoSe sheet. We calculated the exchange constant $J_1$ in decorated ACoSe systems by removing the alkali atoms, but keeping the remaining CoSe sheet unchanged. The $J_1$ in these strained systems is only 70, 85, 65, 52, and 57 K. 
Therefore, the charge transfer between the alkali metal atoms and CoSe layer enhances the RKKY exchange interaction, which emerges as the dominant factor to enhance the ferromagnetism in the CoSe monolayer.
When the adsorbed alkali metal atom ($A$) varies from Li to Cs, the total charge transfer in the $A$CoSe monolayer decreases [see Table S3], leading to a declining trend in the RKKY interaction and $T_{\rm c}$.

Tensile strain can increase the $J_1$ and $T_{\rm c}$ of all the $A$CoSe monolayers, except for LiCoSe [Fig.~\ref{fg4}(d) and Fig.~S14(a, b)]. 
The application of a 4\% tensile strain enhances the $T_{\rm c}$ of NaCoSe, KCoSe, RbCoSe, and CsCoSe to 580 K, 478 K, 407 K, and 342 K, respectively. This enhancement is mainly attributed to the monotonic increase in $J_1$ under tensile strain [Fig. S14(d)]. Conversely, compression suppresses their $T_{\rm c}$.

In contrast, LiCoSe exhibits an opposing strain dependence of $T_{\rm c}$ due to competition between $J_1$ and $J_2$ [Fig. S14(a)]. The bond angles $\theta_1$ and $\theta_2$ evolve monotonically with strain [Fig. S14(c)], and so do the nearest and next-nearest exchange integrals $I_{\rm ex1}$ and $I_{\rm ex2}$.  
However, the Co-Se bond length $d_{\rm Co-Se}$ varies non-monotonically [Fig. S14(b)], modulating Co-3$d$-Se-4$p$ hybridization [Fig. S15] and causing an irregular evolution of the Co magnetic moment $M_{\rm Co}$. The resulting exchange constants, $J = I_{\rm ex} \times M_{\rm Co}^2$, capture these competing effects. 
At 4\% tensile strain, competition between $I_{\rm ex1}$ and $M_{\rm Co}$ slightly increases $J_1$, while $J_2$ drops sharply as both $I_{\rm ex2}$ and $M_{\rm Co}$ decrease, leading to a slight $T_{\rm c}$ reduction. At -4\% compressive strain, bond stretching locally enhances $M_{\rm Co}$, anomalously raising $J_1$ and slightly increasing $T_{\rm c}$.

Finally, we investigated the effect of $U\rm_{eff}$ on the $T_{\rm c}$ of $A$CoSe monolayer. As shown in Fig. S16(a), a non‑zero $U\rm_{eff}$ yields a higher $T_{\rm c}$ than that reported in the manuscript. The strain-dependence of $T_{\rm c}$ remains consistent across different $U\rm_{eff}$ [Fig. S16(b)]. Thus, our main conclusions are robust with respect to the choice of $U\rm_{eff}$.

In summary, using first-principles calculations, we have designed a series of 2D high-$T_{\rm c}$ ferromagnetic metals, the $A$CoSe ($A$ = Li, Na, K, Rb, Cs) monolayers, via alkali-metal adsorption of CoSe monolayer. Notably, LiCoSe exhibits a half-metallic character. Alkali atoms preferentially occupy the hollow sites on top of Se atoms of opposite side of CoSe layer. This adsorption enhances the asymmetry in the occupation of Co-$3d$ orbitals, thereby increasing the magnetic moment of Co ions. Furthermore, charge transfer from the alkali atoms to the CoSe layer strengthens the RKKY-type ferromagnetic coupling, while the lattice distortion weakens the direct antiferromagnetic coupling but enhances the super-exchange interaction between Co ions. Collectively, these factors lead to a large in-plane MAEs (exceeding 833 $\mu$eV/Co) and high Curie temperatures ($T\rm_c > 300 K$). Tensile strain can further enhance the ferromagnetism in $A$CoSe monolayers, except for LiCoSe. With optimal high $T_{\rm c}$ and large MAE, NaCoSe monolayer emerges as the most promising candidate for spintronic devices. This study proposes a strategy for developing practical 2D high-$T_{\rm c}$ ferromagnetic materials in tetragonal lattices for spintronic applications.\newline

See Supplementary Materials for the discussion on the choice of $U_{\rm eff}$; the adsorption of alkali metals; the mechanical, electronic and magnetic properties of $A$CoSe monolayers under different strains.

\begin{acknowledgments}
This work was supported by the National Science Foundation of China (No. 11904313). The numerical calculations have been done in the High Performance Computing Center of Yanshan University.
\end{acknowledgments}

\section*{data avaliability}
The data are available from the authors upon reasonable request.

\bibliographystyle{apsrev4-2}
\bibliography{acose}
\end{document}